# The growth of COVID-19 scientific literature: A forecast analysis of different daily time series in specific settings

Daniel Torres-Salinas[1], Nicolas Robinson-Garcia[2], François van Schalkwyk[3], Gabriela F. Nane[2] and Pedro Castillo-Valdivieso[4]

[1] *torressalinas@go.ugr.es*
Information and Communication Studies department, University of Granada, Granada (Spain)

[2] *elrobinster@gmail.com; g.f.nane@tudelft.nl*
Delft Institute of Applied Mathematics, TU Delft, Delft (Netherlands)

[3] *fbvschalkwyk@sun.ac.za*
DST-NRF Centre of Excellence in Scientometrics and Science, Technology and Innovation Policy, Centre for Research on Evaluation, Science and Technology, Stellenbosch University, Stellenbosch, South Africa

[4] *pacv@ugr.es*
Departamento de Arquitectura y Tecnología de Computadores, University of Granada, Granada (Spain)

## Abstract

We present a forecasting analysis on the growth of scientific literature related to COVID-19 expected for 2021. Considering the paramount scientific and financial efforts made by the research community to find solutions to end the COVID-19 pandemic, an unprecedented volume of scientific outputs is being produced. This questions the capacity of scientists, politicians and citizens to maintain infrastructure, digest content and take scientifically informed decisions. A crucial aspect is to make predictions to prepare for such a large corpus of scientific literature. Here we base our predictions on the ARIMA model and use two different data sources: the Dimensions and World Health Organization COVID-19 databases. These two sources have the particularity of including in the metadata information the date in which papers were indexed. We present global predictions, plus predictions in three specific settings: type of access (Open Access), NLM source (PubMed and PMC), and domain-specific repository (SSRN and MedRxiv). We conclude by discussing our findings.

## Introduction

The average growth in journal articles published is estimated to be at around 3.0% per annum (Johnson et al., 2018) with an increase to 3.9% between 2006 and 2016. The total for developing countries grew more than twice as fast (about 8.6%) (National Science Board, 2018). Unsurprisingly, and given the scale of scientific output, one of the main research topics within the field of scientometrics has been the study of the growth of scientific literature. Indeed, in the 1960s Derek de Solla Price (1963) had already developed a model of the exponential growth of science in what is considered one of the seminal contributions to the field. Although his contribution was not the first attempt to do model growth (e.g., Coles & Eales, 1917; Hulme, 1923), it reflects the predominant role that the study of bibliometric distributions, dynamics of growth and ageing laws of scientific literature has had in the field.

According to Price's model, there are three distinct phases by which literature increases over time. In the first phase there is a slow increment of publications, followed by an exponential increase, and a third phase in which the curve reaches a saturation point. Since then, different studies have tried to refine his approach, by trying to identify the models which can accurately adjust growth curves for the observed increase in scientific literature (i.e., logistic, power or Gumpertz models)[1]. These studies reflect continued efforts to identify models and distributions which can best adjust to different types of scientific literature. Examples of such studies are those conducted by Egghe and Ravichandra (1992) who observe that Social Sciences literature appears to fit a Gompertz-S-shaped distribution, while other literatures follow a power law distribution. Similarly, Zhou (2010) analyses the growth of science in China, while Urbizagástegui and Restrepo (2015) apply exponential models to analyse the Brazilian literature.

---

[1] An overview is provided by Fernandez-Cano et al. (2004).

In this paper we look at scientific growth in exceptional circumstances such as the COVID-19 pandemic. Scientific production on COVID-19 has rocketed in the last year (Torres-Salinas, 2020), reflecting the paramount effort that is being made globally both scientifically and financially to end the global pandemic and to minimize the negative consequences it is having on society. From the scientometric community, efforts have been made to describe the contents of new data sources liberated specifically on the topic of COVID-19 (Colavizza et al., 2020), to compare the coverage of different data sources (Kousha & Thewall, 2020), to analyze the effectiveness of scholarly communication in these pressing times (Homolak et al., 2020; Soltani & Patini, 2020), and its consumption in social media (Colavizza et al., 2020; Thelwall, 2020). The present study is integrated within this stream of literature, building on preliminary findings (Torres-Salinas et al., 2020), and aims to forecast the potential growth of COVID-19 literature to better understand the magnitude of data expected by scientists to cope with the flood of scientific knowledge being produced (Brainard, 2020). We present predictions on the number of COVID-19 publications for 2021. We base our predictions on the Auto-Regressive Moving Average (ARIMA) model and forecast growth in three specific settings. The specific objectives of the paper are summarized as follows:

1. To forecast the growth of publications on COVID-19 in two different databases: Dimensions and WHO.
2. To forecast the growth of publications on COVID-19 in three specific settings to explore the (dis)similarities between them. These are:
   - National Library of Medicine (NLM) databases: Pubmed and PMC
   - Domain-specific scientific repositories: medRxiv and SSRN
   - Type of access to the publications: Open Access and non-Open Access (paywall).

**Material and methods**

We make use of two different databases: Dimensions and World Health Organization (WHO). The former provides a COVID-19-specific dataset named "*Dimensions COVID-19 publications, datasets and clinical trials*" which is available on FigShare. This dataset contains information on four document types: publications, datasets, clinical trials and grants. In this study, we work only with publications, which have a volume of 168,053 records. The second database is the "*COVID-19 global literature on coronavirus disease*", produced by the WHO. In this case we collected metadata for a total of 113,563 records using the export results option that allows for the downloading of the complete database. These two datasets were collected in December 2020. Like Dimensions, the WHO database contains publications from different sources such as international databases (e.g., Pubmed, Elsevier), databases of international organizations (e.g., WHO COVID-19) and repositories (e.g., medRxiv, SSRN, etc.). One of the characteristics of these two specific databases is that they include for each record the exact date on which publications were indexed. In this sense, we have observed a two-day delay in the indexing dates for the WHO database with respect to Dimensions. This information allows us to establish the daily growth in the number of publications. Table 1 presents a summary of the main characteristics of both databases.

Three different datasets were generated for each database, producing a total of eight time series (Table 2). The first two time series account for the total number of records per day in each database. Two additional time series include the number of published Open Access (OA) and non-OA documents per day. The last four time series refer to the number of documents published by repository. We report predictions of growth for the following repositories: PubMed, PMC, medRxiv and SSRN.

**Table 1. Main characteristics of the analysed databases: Dimensions and WHO**

|  | **Dimensions** | **WHO** |
|---|---|---|
| Link | https://tinyurl.com/y3bhurmm | https://tinyurl.com/rdkr4c7 |
| Last download | 6 December 2020 | 5 December 2020 |
| Starting day | >1 January 2020 | 7 April 2020 |
| End day | 16 November 2020 | 6 December 2020 |
| Type of publications | article, preprint, chapter, book monograph, preprint and proceedings | article, monograph, non-conventional and preprint |
| Fields and information provided | Bibliographic description Record provider Citations Altmetrics Open Access information | Bibliographic description Record provider |
| No. of records | 168.053 | 118.200 |
| No. of information sources | 43 | 24 |
| Main type and number of information sources | International Databases (2) Repositories (41) | International Databases (2) Repositories (10) Internal Databases (2) Others (10) |
| Main information sources and percentage of total records | Pubmed (47%) PMC (36%) medRxiv (4%) SSRN (4%) | Pubmed (51%) Internal database (30%) Elsevier (7%) medRxiv (6%) |

**Table 2. Contexts & scenarios: general view of the different timelines established**

| Dataset | Time series name | Subseries an coverage periods) | Database | Forecast Starting and ending date |
|---|---|---|---|---|
| General | TS1-General | TS1a - Total documents per day in WHO TS1b - Total documents per day in Dimensions | WHO Dimensions | 07/11/2020 - 06/11/2021 14/10/2020 - 13/10/2021 |
| Open Access | TS2- Access | TS2a - Total Open Access documents per day TS2b - Total Non-Open Access documents per day | " " | " " |
| Sources | TS3-Sources | TS3a - Total documents per day in Pubmed TS3b - Total documents per day in PMC TS3c - Total documents per day in meRxiv TS3d - Total documents per day in SSRN | " " " " | " " " " |

The prediction of publication growth requires adequate tools to analyze historical data. There are several types of models that can be used for time-series forecasting. In this study we make use of ARIMA, which is a widely used forecasting method (Hyndman & Athanasopoulos,

2018). In the ARIMA model, only historical data of the variable of interest are used and forecasts are modelled as a linear combination of past observations and past error terms of the model (Hyndman & Khandakar, 2008). An ARIMA model is characterized by three parameters $(p, d, q)$ where:

-   $p$ refers to the number of past values accounted in the model,
-   $d$ indicates the order of difference for attaining stationarity, and
-   $q$ specifies the number of error terms included in the model.

The ARIMA model can be used for non-stationary data, that is, for data in which the average and variance change over time. Since all eight time series exhibit a trend, the data are non-stationary and ARIMA handles non-stationarity by differencing subsequent observations. The necessary number of differencing to ensure stationarity is indicated by the parameter $d$. The three parameters are estimated from data, usually via maximum likelihood estimation (Hyndman & Athanasopoulos, 2018).

ARIMA models were fitted to the eight time series included in Table 2. All the analyses were conducted on an Ubuntu 18.04.1 machine, with R version 3.6.3 and RStudio version 1.1.456. The forecast analysis was carried out with a one-year window and specific results are offered for three-month windows. Along with point estimates, a 95% confidence interval accounts for the forecast uncertainty. Datasets and analyses of this study are openly accessible at https://doi.org/10.5281/zenodo.4478251.

**Results**

*Evolution of COVID-19 scientific literature*

The cumulated number of publications in Dimensions and WHO are presented in Figure 1. Dimensions indexed a total of 168,053 records and WHO a total of 118,200. As reported in Table 1, there are differences in the coverage of each source; while Dimensions covers records published in the last 10 months, WHO only does so for the last 8 months. Along with differences in size and period covered, we observe differences in the growth rate. In the case of Dimensions, it is more pronounced, especially from June onward. Both general time series fit a linear model, with $R^2$ values above 0.9 ($R^2 = 0.931$ in WHO; $R^2 = 0.851$ in Dimensions).

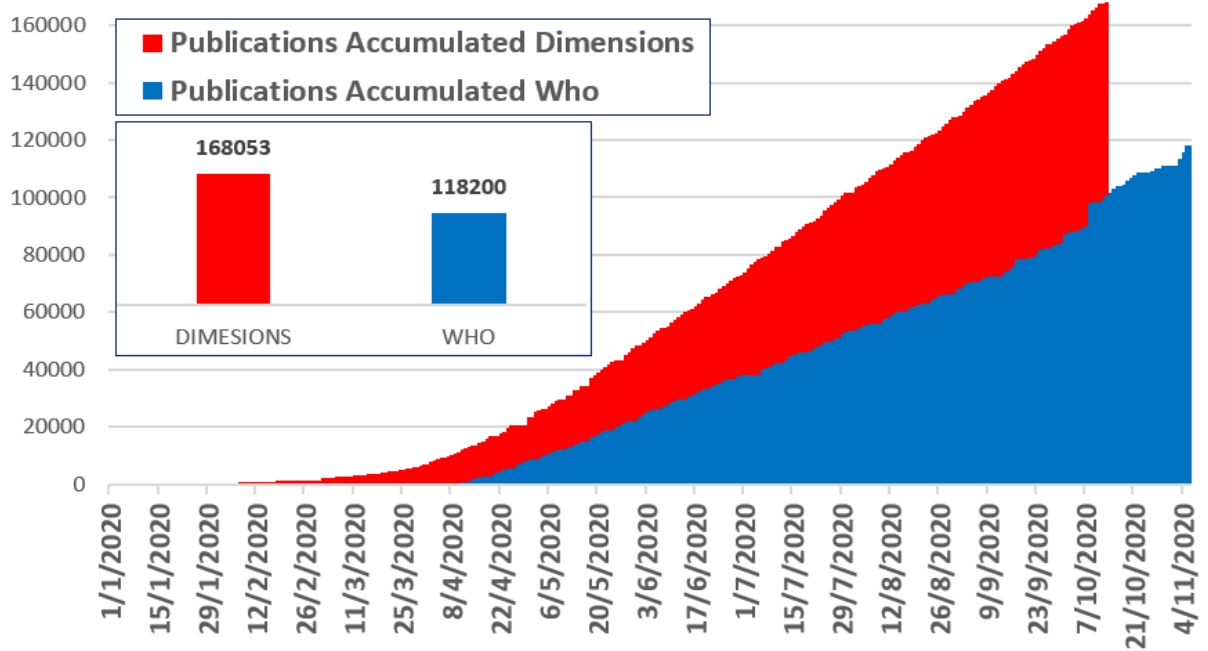

**Figure 1. Accumulated number of records in Dimensions and WHO**

Figure 2 shows the results for six time series. Figure 2A shows the results for Pubmed and PMC. These two repositories are the most prevalent sources in the Dimensions dataset, with PubMed alone including 47% of the share in this database (78,841 records). Figure 2B shows the time trend for medRxiv and SSRN. In this case we observe that both sources have similar volumes (7,002 and 6,002 records respectively) and a similar growth trend, with exponential growth until June 2020. Finally, Figure 2C compares the time series of OA and non-OA publications. Here the differences both in size and growth trends are very significant. OA literature is approximately five times larger than the non-OA and follows an exponential trend. In comparison, the growth of the non-OA publications is low.

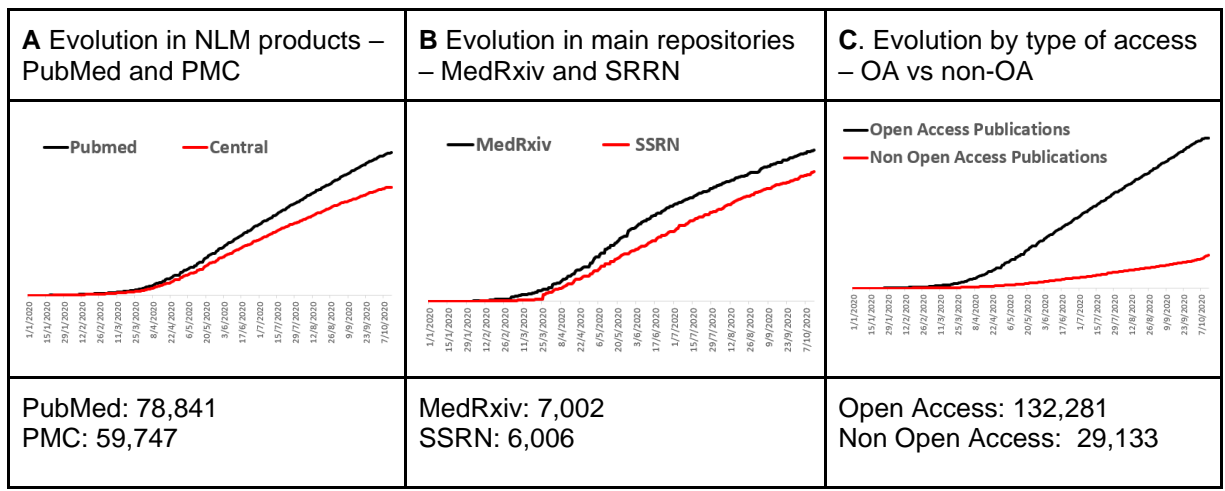

**Figure 2. Time trend on the accumulated number of records in NLM databases, main repositories, and open access (OA)**

*Forecasting*

Figure 3 and Figure 4 present the predictions for the Dimensions and WHO time series. We include our predictions along their uncertainty bounds. As observed, the lower bound shows a deceleration of growth, while in the two other cases it reflects a sustained rate of growth over time.

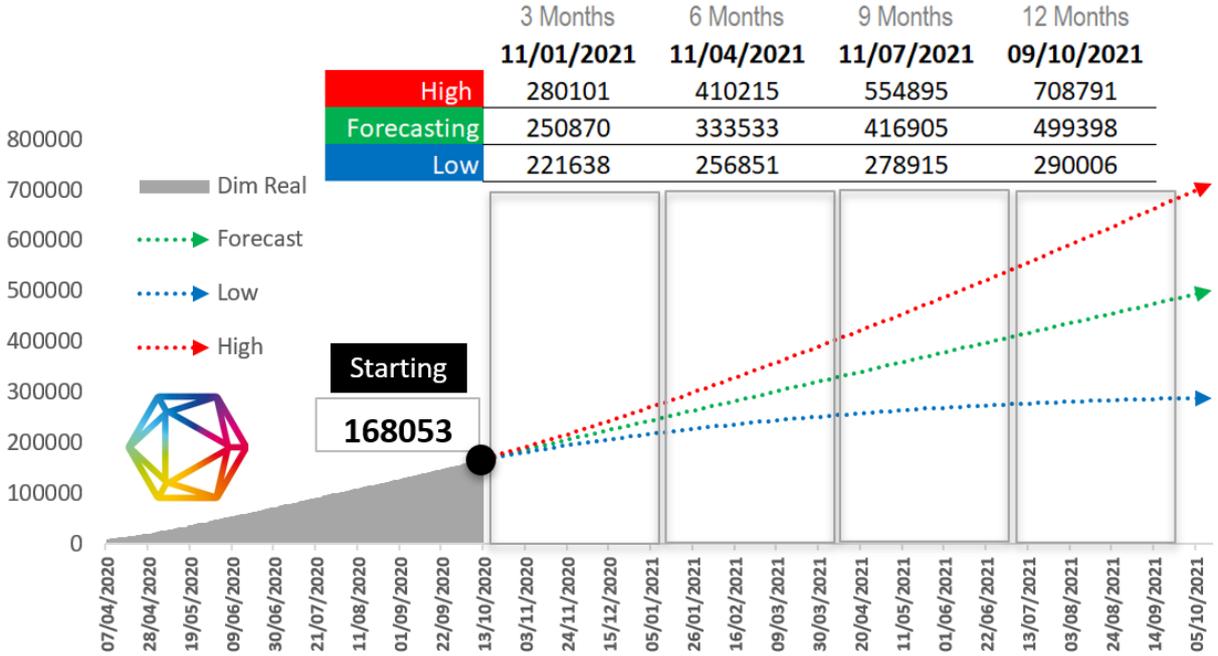

| | 3 Months | 6 Months | 9 Months | 12 Months |
|---|---|---|---|---|
| | **11/01/2021** | **11/04/2021** | **11/07/2021** | **09/10/2021** |
| **High** | 280101 | 410215 | 554895 | 708791 |
| **Forecasting** | 250870 | 333533 | 416905 | 499398 |
| **Low** | 221638 | 256851 | 278915 | 290006 |

**Figure 3. Forecasted growth of overall publications in Dimensions for 2021. Predicted growth (green) and upper (red) and lower bounds (blue) accounting for a 95% uncertainty interval. Forecasts are provided every three months**

According to the ARIMA model, the forecast is that by the beginning of October 2021, the number of COVID-19 publications will reach half a million (499,398) according to Dimensions, with an upper bound of 708,791 records. This means that we expect the volume of COVID-19 publications to double by June 14th, 2021. If we consider the upper bound, the number of publications will double by February 20th, 2021.

A similar growth trend is observed for publications in the WHO database (Figure 4); the forecast is that 389,418 publications will be reached by the beginning of November 2021. The most likely maximum number of publications that is expected to be reached in the WHO database is 559,404. Based on the total number of records included on the date the data was collected, we should expect this number to double on June 11th, 2021. If we consider the upper bound of the forecast, the number of publications will double on February 24th, 2021 with 236,282. In both cases, the dates of growth and figures are similar, with Dimensions doubling the number of records in 7.8 months (243 days) and the WHO database in 7.13 months (217 days).

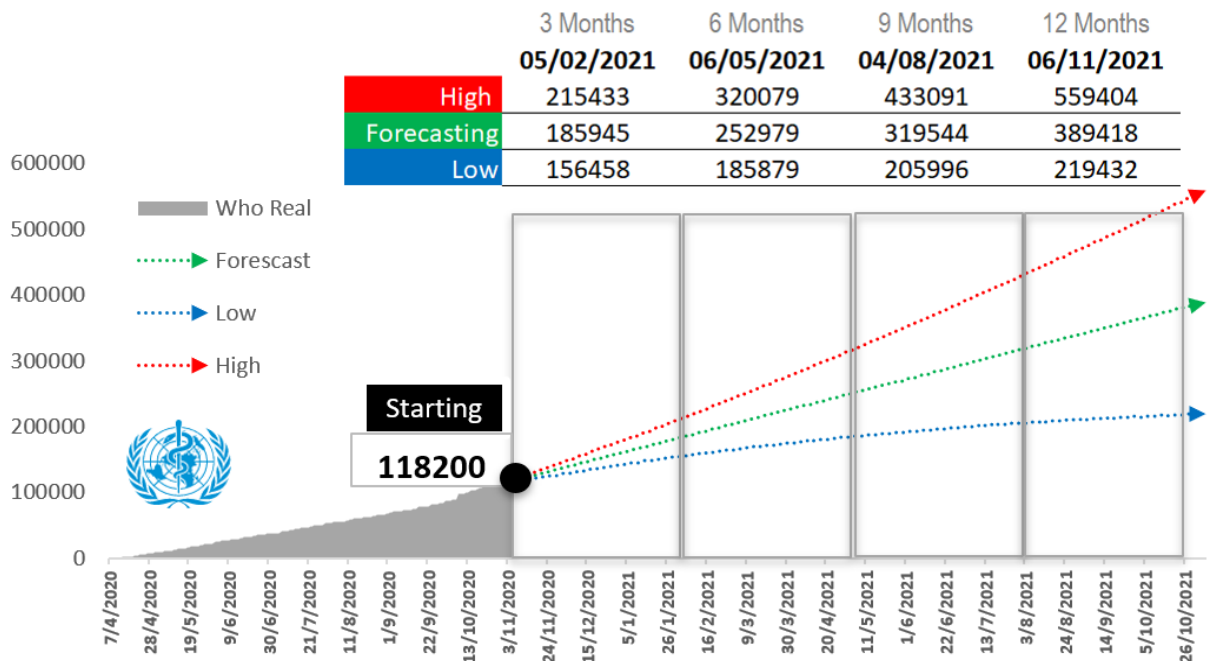

**Figure 4. Forecasted growth of overall publications in the WHO for 2021. Predicted growth (green) and upper (red) and lower bounds (blue) accounting for a 95% uncertainty interval. Forecasts are provided every three months**

*Publication settings*

Table 3 complements the general predictions in Dimensions and the WHO databases. The data is disaggregated and filtered based on three different settings: 1) type of access, 2) NLM source, and 3) domain-specific repository.

There are a total of 132,281 OA publications in the dataset (Table 3A). We observe an increase of 40% in their volume by the 14th of September, 2021. But the most intriguing growth is that of non-OA publications. Starting at an initial size of 29,133 at the time of the data retrieval, we expect an increase by a factor of 3.7 in the six following months, and 6.2 a year later. This spectacular increase is given by the rapid increase during the last period of registered data, as observed in Figure 2C. The upper growth scenario multiplies the starting non-OA papers by a factor of almost 11.

Similar forecast growth estimates are registered for PubMed and PMC (Table 3B). We estimate both sources will double their number of publications in a year. These two databases currently have a significant number of documents indexed, thus the effort required to double their size. Table 3C shows that these time windows are shorter for the two repositories analyzed, probably due to their smaller size. In the case of medRxiv, we estimate that the number of COVID-19 publications will increase by a factor of 15 in the next six months, and by a factor of 19 in a year (from 7,004 publications to 133,328). For SSRN, a more pronounced growth rate is estimated. In six months, the number of publications is expected to multiply by a factor of 17 and in twelve months by a factor of 25 (from 6,008 publications to 151,185).

**Table 3. Forecast growth of publications by case scenario: A) type of access, B) NLM source and C) domain-specific repositories. It includes the predicted value and the upper bound of a 95% uncertainty level. Predictions are provided every three months**

| **A** *Time series by type of access (Open Access vs. non-Open Access)* | | | | | |
|---|---|---|---|---|---|
| Type | **Starting** | **3 Months** | **6 Months** | **9 Months** | **12 Months** |
| | 13/10/2020 | 11/01/2021 | 11/04/2021 | 11/07/2021 | 14/09/2021 |
| OA | 132,281 | **155,661** | **176,705** | **197,983** | **219,027** |
| | | High: 191,926 | High: 281,168 | High: 392,178 | High: 518,526 |
| Non-OA | 29,133 | **81,482** | **106,952** | **146,236** | **185,089** |
| | | High: 10,6783 | High: 151,899 | High: 228,054 | High: 309,963 |
| **B** *Time series by NLM data source (PubMed vs. PMC)* | | | | | |
| Database | **Starting** | **3 Months** | **6 Months** | **9 Months** | **12 Months** |
| | 13/10/2020 | 11/01/2021 | 11/04/2021 | 11/07/2021 | 14/09/2021 |
| PubMed | 78.841 | **98,879** | **118,236** | **137,808** | **158,025** |
| | | High: 116,539 | High: 168,792 | High: 231,599 | High: 304,949 |
| PMC | 59.744 | **74,644** | **89,321** | **104,162** | **119,492** |
| | | High: 89,282 | High: 129,123 | High: 176,706 | High: 232,105 |
| **C** *Time series by domain-specific repository (MedRxiv vs. SSRN)* | | | | | |
| Repository | **Starting** | **3 Months** | **6 Months** | **9 Months** | **12 Months** |
| | 13/10/2020 | 11/01/2021 | 11/04/2021 | 11/07/2021 | 14/09/2021 |
| MedRxiv | 7.004 | **8,589** | **10,140** | **11,708** | **13,328** |
| | | High: 10,849 | High: 16,618 | High: 23,735 | High: 32,174 |
| SSRN | 6.008 | **8,259** | **10,525** | **12,817** | **15,185** |
| | | High: 10,186 | High: 15,731 | High: 22,284 | High: 29,863 |

## Discussion and concluding remarks

In this paper we present a forecasting analysis on the production of COVID-19-related scientific literature for 2021. We contribute to existing literature analysing the growth of science, a topic of interest since the very inception of scientometrics, with the pioneering works of Derek de Solla Price. However, we focus on a very particular type of scientific literature, that is, publications related to the COVID-19 pandemic. The scientific communication system has never generated as much interest, both scientific and societal, as it is generating during the COVID-19 crisis (Zastrow, 2020). Our results point towards potential scenarios for which infrastructure, communication strategies and policy actions must be coordinated to maximize the result of such paramount scientific effort (Brainard, 2020). We use the ARIMA model to predict literature growth as, despite the simplicity of this model, it proved to be highly accurate in our preliminary findings (Torres-Salinas et al., 2020). In times of social mistrust and fake news (Lazer et al., 2018), the production of new scientific knowledge must be accompanied by effective science communication strategies. The emergence of sources such as the WHO database and the CORD19 dataset already reflect a contribution to such efforts.

Although there is still debate as to what constitutes COVID-19-related literature (Kousha & Thelwall, 2020), the two databases have the unique feature of indicating daily indexing dates, which helps modelling data for predicting growth. Also, the level of transparency of these sources allows one to determine potential misrepresentations in certain fields (e.g., the inclusion of SocArxiv shows promise as to having a good coverage of social science fields). Our analysis by scenario points towards different levels (and, potentially, models) of growth depending on the data source used. Further steps will require looking into differences in growth rate by fields as well as considering external socio-economic and health factors which may affect the growth of scientific literature on this research front.

The urgency of the extraordinary health and financial crisis triggered by the pandemic has pushed the expansion of Open Acess and the inclusion of preprints as tacitly accepted scientific publications (although with many cautionary notes). This presents further challenges related to the control of scientific quality, certainty and rigour, although it is still too early to tell whether quality is being compromised in these pressing times of accelerated scientific discovery (Abritis, Marcus & Oransky, 2020). The fact that science is squarely in the social spotlight makes it especially vulnerable when errors are committed or when messages are misinterpreted. In the light of this framing, we believe that further research on this matter should continue to further our understanding of the growth not only of scientific publications, but also of the social reaction to science, and of the types of access by which scientific publications are made available.

## Acknowledgments
This study is part of the project 'Scientific communication in times of Corona virus' funded by the TU Delft COVID-19 Response Fund.